\documentclass[aps,prb,twocolumn,amsmath,amssymb,showpacs]{revtex4-1}
\usepackage{framed}
\usepackage{mathtools}
\usepackage{graphicx}
\usepackage{dcolumn}
\usepackage{amsmath}
\usepackage{amssymb}
\usepackage{subfigure,amsmath,verbatim,moreverb,bm}
\usepackage{color}

\def\be{\begin{equation}}
\def\ee{\end{equation}}
\def\ber{\begin{eqnarray}}
\def\eer{\end{eqnarray}}

\def\pv{{\bf p}}

\def\nn{\nonumber}

\def\Acal{{\bf {\cal A}}}
\begin{document}
\title{Ballistic spin transport in the presence of interfaces with strong spin-orbit coupling}
\author{J. Borge$^1$}
\author{I. V. Tokatly$^{1,2}$}
\affiliation{$^1$Nano-Bio Spectroscopy group, Departamento de F\`isica de Materiales,
Universidad del Pa\`is Vasco UPV/EHU, E-20018 San Sebasti\`an, Spain}
\affiliation{$^2$IKERBASQUE, Basque Foundation for Science, E-48011 Bilbao, Spain}

\begin{abstract}
The inversion symmetry breaking at the interface between different materials generates a strong interfacial spin-orbit
coupling (ISOC) that may influence the spin and charge transport in hybrid structures. Here we use a simple analytically solvable model to study in the ballistic approximation various spin transport phenomena induced by ISOC in a bilayer metallic system. In this model a non-equilibrium steady state carrying a spin current is created by applying a spin dependent bias across the metallic junction. Physical observables are then calculated using the scattering matrix approach. In particular we calculate the absorption of the spin current at the interface (the interface spin-loss) and study the interface spin-to-charge conversion. The latter consists of an in-plane interface charge current generated by the spin dependent bias applied to the junction, which can be viewed as a spin galvanic effect mediated by ISOC. Finally we demonstrate that ISOC leads to an interfacial spin current swapping, that is, the ``primary'' spin current flowing through the spin-orbit active interface is necessarily accompanied with a ``secondary'' swapped spin current flowing along the interface and polarized in the direction perpendicular to that of the ``primary'' current. Using the exact spin continuity equation we relate the swapping effect to the intefacial spin-loss, and argue that this effect is generic and independent on the ballistic approximation used for specific calculations.
\end{abstract}
\maketitle
\section{Introduction}
During the last 15 years spin-orbit coupling (SOC) has emerged as one of the most popular topics in
spintronics \cite{Zutic04,Sinova2015, Hirsch99, Zhang00, Murakami03, Sinova04, Handbook,
Raimondi_2Dgas_PRB06, Culcer_SteadyState_PRB07, Culcer_Generation_PRL07, Culcer_SideJump_PRB10, TsePRB05,
GalitskiPRB06, Tanaka_NJP09,Hankiewicz09,TenYears2010}. The coupling between the spin and charge degrees of
freedom  permits the manipulation of the magnetic properties through electric fields. This fact has lead to
the study of many different effects produced by SOC in non-magnetic materials.
The spin Hall effect (SHE), which is the appearance of a spin current produced by an external electric
field/current perpendicular to it, has focused the attention of the field at its beginning
\cite{Dyakonov71,Kato04, Sih05, Wunderlich05, Stern06, Stern08,Raimondi05}.
The Edelstein effect (EE), also known as the current induced spin polarization, or the inverse spin galvanic effect, consists, instead, in the appearance of a spin polarization in response to a perpendicular
charge current, and it has been proposed as a
promising way of achieving all-electrical control of magnetic
properties in electronic circuits
\cite{Lyanda-Geller89,Edelstein1990,Ioan2010,Kato04,Sih05,Inoue2003,Yang2006,Chang2007,Koehl2009,Kuhlen2012}.
Their inverses, i.e. the inverse spin Hall (ISHE) and the inverse Edelstein effects (IEE), have also been
studied theoretically and
found experimentally \cite{Valenzuela_Nat06,Takahashi_Revese_PRL07,Takahashi_GSH_NatMater08,Shen2014,Rojas2013}.
While these effects rely on the spin-to-charge conversion, the SOC can be also responsible of spin-to-spin conversion.
The example is a spin swapping effect (SSE) that consists in the appearance of a ``secondary'' spin current as a response to a
non-equilibrium ''primary'' spin current in such a way that the directions of flow and the spin in the secondary current are ``swapped'' with respect to those in the primary spin flow \cite{Lifshits2009,Shen04,Saidaoui2016}. Observation of the effect presents a
difficult issue, as it requires to produce and quantify a spin current that acts like a source.
Sometimes it is complicated to distinguish
between the SSE and the so-called spin Hanle effect, where the secondary current is produced by the combination
of the external electric field, the ferromagnetic exchange field, and the SOC \cite{Shen2015}.
If we are dealing with pure non-magnetic materials this duality vanish.

In bulk materials the effects of SOC are usually weak. However, the inversion symmetry breaking across the interface of two different materials may produce a giant SOC \cite{Ast2007,Mathias2010,Rybkin2010,Moreschini2009,Shikin2013}.
The strong interfacial SOC (ISOC) has attracted a lot of attention in the field of hybrid magnetic/non-magnetic structures
because of its relevance for many spintronics effects like spin-orbit torques \cite{Miron2011,Amin16,Wang2016}.

When a spin current passes through an interface with ISOC a part of
this spin current is absorbed. This effect is called spin loss and it is crucial to take it into account in order to interpret and quantify spin pumping
experiments \cite{Rojas2014,Chen_pumping_2015}.
The effect of ISOC in non-magnetic/non-magnetic junctions has been actively studied in the last few years. The efficiency of the interfacial spin-to-charge converting has been demonstrated in various, metal/metal, \cite{Rojas2013,Isasa2016}  metal/insulator
\cite{Karube2015,Karube2016}, and in topological insulator interface \cite{Shiomi2014} experiments, using
spin pumping and spin lateral valves techniques.
Theoretically, metal/insulator transport phenomena induced by ISOC in non-magnetic materials has been studied in thin
metal films \cite{Wang2013,Borge2014} and semi-infinite junction geometries \cite{Tokatly2015}.

In lateral spin valves experiments a non-equilibrium spin distribution is  generated by passing a current through a
ferromagnet \cite{Morota2011,Niimi2012,Isasa2016,Sagasta2016}. This spin distribution is transported in a material
with low SOC, such as Cu or Al, to a material with a strong SOC at the interface where a part of the incoming spin current is absorbed and/or converted to a charge current parallel to the interface. In general one can suggest at least three different mechanisms for the generation of the charge current in this kind of experimental setup. These are (i) the ISHE in the bulk of materials forming the junction, (ii) the IEE in the interface/surface Rashba-splitted band, and (iii) the spin-to-charge conversion due to the interference of the spinor Bloch states scattered off the interface with strong ISOC. The first two mechanisms as well as their relative contribution to the generation of the charge current have been extensively discussed in the literature \cite{Morota2011,Niimi2012,Isasa2016,Sagasta2016,Rojas2013,Karube2015,Karube2016}. To the best of our knowledge the last, ``interference channel'' of the spin-to-charge conversion have not been studied in the context of the spin transport through a metal/metal
interface. However, by analogy with the current
induced surface spin accumulation studied in Ref.~\onlinecite{Tokatly2015} one can expect that this channel may give a contribution comparable to that of the IEE in the interface band. In fact, it has been shown in \cite{Tokatly2015} that at lest for some materials the contribution of the bulk scattering states to the surface spin accumulation dominates over the contribution produced via the EE in the Rashba-splitted surface band. Investigation  of the spin-dependent interference mechanism of the interface spin-to-charge and spin-to-spin conversion is the main subject of the present work.


Specifically in this paper we study spin transport across a planar metal/metal interface supporting ISOC, within a scattering matrix approach, in the ballistic limit. To make the physics as clear as possible we exclude the bulk ISHE and a possible presence of interface bands by modeling the junction as two semi-infinite Fermi gases with a potential step due to different work functions. The SOC appears as the derivative the potential, which generates a Rashba-like SOC term localized at the interface. Within the ballistic approximation we adopt the Landauer approach and generate a pure spin current by applying a spin bias that is modeled as difference of spin-dependent chemical potentials across the interface. Using this model we focus on three different phenomena, the spin loss at the interface, the conversion of the incident spin current to the interface charge current, and the interface spin swapping effect, i.e., the spin-to-spin conversion mediated by the ISOC.
%
In a forthcoming paper \cite{Borge2017}  we study these interface at the quantitative level in a realistic tungsten-aluminium junction using first principle, DFT-based transport theory methods.

\begin{figure}
\begin{center}
\includegraphics[width=3.5in]{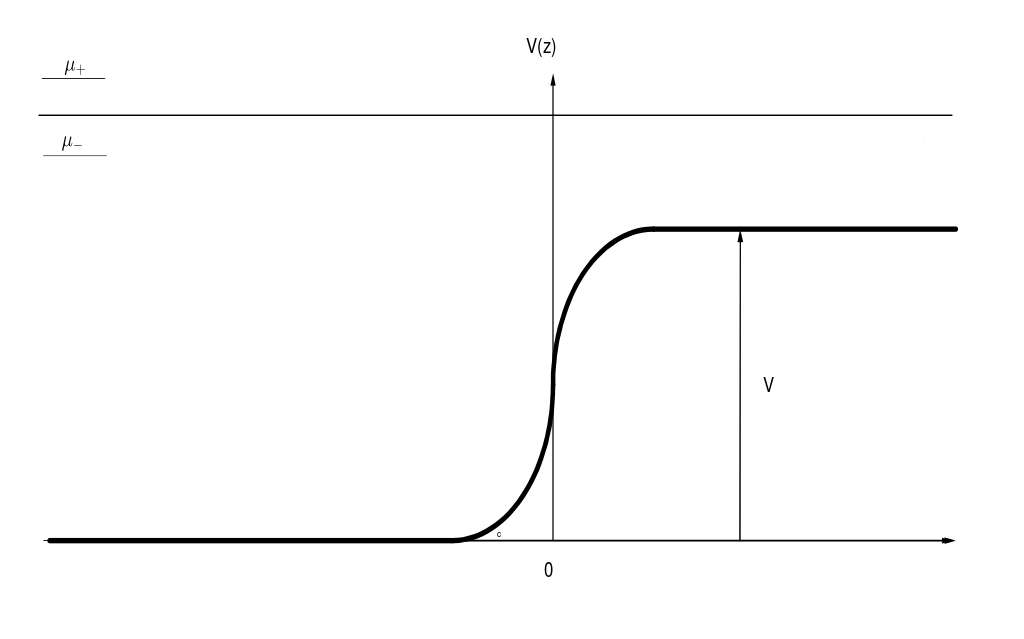}
\caption{ Scheme of the potential of the metal-metal junction.
$\mu_{\pm}$ are the chemical potentials for $\pm$spins species in the x-axis}
\label{Chem}
\end{center}
\end{figure}

This paper is organized as follows.  In Sec.~\ref{sec_model} we introduce and discuss the model.
In Sec.~\ref{sec_Loss} we study the spin loss effect, that is, the absorption of the spin current at the interface produced by the ISOC.
In Sec.~\ref{sec_StC} we show that the spin-dependent scattering at the interface with ISOC leads to the spin-to-charge conversion. In particular we calculate the interfacial charge current that is produced by the spin bias (difference of spin-dependent chemical potentials) applied to the junction. This effect can be viewed as an interfacial analog of the ISHE, which originates solely from the scattering states. In Sec.~\ref{sec_Swappin} we first present general arguments based on the spin continuity equation, which show that the spin loss at the interface implies the existence of the intefacial spin current swapping. Then we explicitly calculate in our model the spin-to-spin conversion due to ISOC.
Sec.~\ref{sec_conclusions} presents our summary and conclusions.

\section{The model}
\label{sec_model}
Our aim is to describe a ballistic spin transport across the interface between two different metals. To simplify calculations and emphasize the physical picture we consider the simplest possible ``minimal'' model that captures the most important physical aspects of the effects of interest. We model our system as a free electron gas (jellium model) in the presence of a potential forming a
step at the interface located at $z=0$ (see Fig.~\ref{Chem}). The potential step represents different work functions of the two metals. The system is assumed to be translation invariant in the x-y plane.
The presence of the interface (the potential step) breaks the inversion symmetry and generates a local SOC term which is responsible for the transport phenomena that we will study later on. The model is described by the following Hamiltonian
\begin{equation}
\label{H1}
H=\frac{p^2}{2m}-\frac{\partial_z^2}{2m}+V(z)+\gamma V'(z)({\bf\hat{z}}\times{\bf p})\cdot{\bm\sigma},
\end{equation}
where ${\bf p}$ is the two-dimensional momentum in the x-y plane, $V(z)$ the potential step, ${\bm\sigma}$ is a vector of Pauli matrices, and $\gamma$ is a material dependent parameter which describes the strength of SOC at the interface.
The first two terms of the Hamiltonian describe the kinetic energy of the electrons, the third one represents
the potential step due to the different work functions, and the fourth term localized at the interface corresponds to SOC due
to the gradient of this potential barrier, the ISOC.

Because of the translation invariance in the x-y plane the eigenfunctions read
\begin{equation}
\psi_{{\bf p},k}({\bf r},z)=e^{i{\bf p}\cdot {\bm\rho}}\varphi_k(z),
\end{equation}
where ${\bm\rho}=(x,y)$ is the in-plane coordinate, and $\varphi_k(z)$ are spinor scattering states in $z$-direction which are labeled by the wave vector $k$ of the incoming wave. In the complete set of the scattering states we distinguish two orthogonal subsets of eigenfunctions: (i) the states $\overrightarrow{\varphi}$ incoming from the left, and (ii) the states $\overleftarrow{\varphi}$ incoming from the right. Away from the interface the wave functions corresponding to the energy $\epsilon=(\pv^2+k^2)/2m$ have the following form
\begin{eqnarray}
\nn
\overrightarrow{\varphi}_{k\sigma}(z)&=&\left\{\begin{array}{c}\hspace{0.01cm}(e^{ikz}+\hat{r}_ke^{-ikz})
\chi_{\sigma}\\\hspace{1cm}
\hat{t}_ke^{ik'z}\chi_{\sigma}\end{array}\begin{array}{c}\hspace{0.75cm}if\hspace{0.7cm}z<0\\\hspace{0.8cm}if\hspace{0.75cm}z>0\end{array}\right. \\
\label{states}
\overleftarrow{\varphi}_{k\sigma}(z)&=&\left\{\begin{array}{c}\hspace{0.01cm}(e^{-ik'z}+\hat{r}'_ke^{ik'z})
\chi_{\sigma}\\\hspace{1.1cm}
\hat{t}'_ke^{-ikz}\chi_{\sigma}\end{array}\begin{array}{c}\hspace{0.45cm}if\hspace{0.65cm}z>0\\\hspace{0.45cm}if\hspace{0.65cm}z<0\end{array}\right.
\end{eqnarray}
where $k'=\sqrt{k^2-2mV}$, $\hat{r}_k$ and $\hat{r}'_k$ are the $2\times 2$ matrix reflexion coefficients,
$\hat{t}_k$ and
$\hat{t}'_k$ are the matrix transmission coefficients, and the spinor factors $\chi_{\sigma}$ with $\sigma=\pm$ are the basis vectors spanning the spinor subspace. In the following, for definiteness, we choose $\chi_{\sigma}$ to be the eigenfunctions of $\sigma_x$, which read
$\chi_{\pm}^{\dagger}=(1,\pm 1)/\sqrt{2}$.
Apparently the expressions of Eq.~(\ref{states}) are only valid when $k^2>2mV$, that is, when the energy of the scattering states is larger then the height of the interface potential barrier. It can be easily proven that the electrons with $k^2<2mV$ do not contribute to the transport effects we are considering in this work.

In general, the matrix scattering coefficients can be represented as follows
\begin{eqnarray}
 \label{r-general}
&& \hat{r}_k=r_0\sigma_0+{\bf r}\cdot\pmb{\sigma}, \quad \hat{r}'_k=r'_0\sigma_0+{\bf r'}\cdot\pmb{\sigma},\\
\label{t-general}
&& \hat{t}_k=t_0\sigma_0+{\bf t}\cdot\pmb{\sigma}, \quad \hat{t}'_k=t'_0\sigma_0+{\bf t'}\cdot\pmb{\sigma},
\end{eqnarray}
where, $\sigma_0$ is a $2\times 2$ unit matrix, and vectors ${\bf r}$, ${\bf r'}$, ${\bf t}$, and ${\bf t'}$ describe the spin dependent (spin flip) part of the scattering at the interface. A nontrivial spin dependent part of the scattering coefficients, appearing due to the ISOC, is the physical origin of the spin-to-charge and spin-to-spin conversion effects that will be considered in the next sections.

It is convenient to define the following dimensionless parameters, $\nu=\sqrt{2mV}/k_F$ and
$s=\gamma k^2_F$, which, respectively, quantify the difference of the work functions, and the strength of ISOC in the units of Fermi energy $E_F=k_F^2/2m$ of the left metal. In the physical situation we are describing here the SOC and the potential barrier
are smaller than $E_F$ which implies $\nu,s<1$.

In order to have closed analytic expressions for the scattering coefficients in Eqs.~(\ref{states})-(\ref{t-general}) we will choose the potential to be $V(z)=V\Theta(z)$, which implies $V'(z)=V\delta(z)$.
%
By solving explicitly the scattering problem for the Hamiltonian of Eq.~(\ref{H1}) with $V(z)=V\Theta(z)$ we find
\begin{eqnarray}
\nn
t_0&=&\frac{2k}{k+k'}\frac{1}{1+\frac{(\nu^2 s p)^2}{(k+k')^2}},\hspace{1cm}
t_0'=\frac{2k'}{k+k'}\frac{1}{1+\frac{(\nu^2 s p)^2}{(k+k')^2}},\\
\nn
r_0&=&-1+t_0=\frac{(k-k')-\frac{(\nu^2 s p)^2}{(k+k')}}{k+k'}\frac{1}{1+\frac{(\nu^2 s p)^2}{(k+k')^2}},\\
\nn
r_0'&=&-1+t_0'=\frac{(k'-k)-\frac{(\nu^2 s p)^2}{(k+k')}}{k+k'}\frac{1}{1+\frac{(\nu^2 s p)^2}{(k+k')^2}},\\
\nn
{\bf t}&=&{\bf r}=i(\hat{\bf z}\times{\bf p})\frac{s\nu^2}{k+k'}t_0\\ \label{Scat2}
{\bf t}'&=&{\bf r}'=i(\hat{\bf z}\times{\bf p})\frac{s\nu^2}{k+k'}t'_0.
\end{eqnarray}
The scattering coefficients have some general properties which will be useful in the next sections:
\begin{eqnarray}
\label{r0}
r_0^{*}{\bf r}+{\bf r}^{*}r_0&=&r_0^{'*}{\bf r'}+{\bf r}^{'*}r'_0=0\\
\label{t0}
t_0^{*}{\bf t}+{\bf t}^{*}t_0&=&t_0^{'*}{\bf t'}+{\bf t}^{'*}t'_0=0
\end{eqnarray}
In addition the charge current conservation implies the following identity
\be
\label{current-conserv}
\frac{k}{m}\left(1-r_0^*r_0-{\bf r}^*{\bf r}\right)= \frac{k'}{m}\left(t_0^*t_0+{\bf t}^*{\bf t}\right)
\ee

Equations (\ref{states})-(\ref{Scat2}) define the complete and orthonormal set of scattering states for our model.
In the next sections we will use these scattering states to study a steady state ballistic spin transport across the interface with ISOC. To create a nonequilibrium density matrix supporting steady spin and/or charge flows we populate the states $\overrightarrow{\varphi}_{\sigma}$ incoming from the left metal with the chemical potential $\mu_{L\sigma}$,  and the states $\overleftarrow{\varphi}_{\sigma}$ incoming from the right metal with the chemical potential $\mu_{R\sigma}\ne\mu_{L\sigma}$. In this paper we are interested in the situation of a pure spin current which corresponds to $\sum_{\sigma}\mu_{R\sigma}=\sum_{\sigma}\mu_{L\sigma}$.

\section{Spin loss at the interface}
\label{sec_Loss}
In this Section we calculate the spin loss which occurs when a spin current passes through a spin-orbit active interface.
We will consider the case shown in Fig.~\ref{Chem}, which corresponds to different
spin-plus and spin-minus chemical potentials ($x$-polarized spin bias) in the left metal ($\mu_{L+}\ne\mu_{L-}$ with $\mu_{L+}+\mu_{L-}=2E_F$), and $\mu_{R+}= \mu_{R-}=E_F$ in the right metal. As we will see in the following, the effect of different spin-chemical potential in  the right metal can be calculated straightforwardly in the same way. The total spin current flowing along the $z$-direction and polarized along $x$-axis, $J_z^x$, is given by the following formula
\be
\label{J1}
J_z^x=\sum_{{\bf p},k,\sigma}f_F\left(\epsilon_{k,p}-\mu_{L\sigma}\right)j_{zL}^{x\sigma}
+f_F\left(\epsilon_{k,p}-\mu_{R\sigma}\right)j_{zR}^{x\sigma},
\ee
where $f_F(E)$ is the Fermi distribution function,  $\epsilon_{k,p}=(k^2+p^2)/2m$, and
\begin{eqnarray}
\nn
j_{z,L}^{x,\sigma}(z)&=&-\frac{i}{4m}\left(\overrightarrow{\varphi}_{\sigma}^{\dagger}(z)\sigma_x\partial_z\overrightarrow{\varphi}_{\sigma}(z)-(\partial_z\overrightarrow{\varphi}_{\sigma}^{\dagger}(z))\sigma_x\overrightarrow{\varphi}_{\sigma}(z)\right)\\
\nn
j_{z,R}^{x,\sigma}(z)&=&-\frac{i}{4m}\left(\overleftarrow{\varphi}_{\sigma}^{\dagger}(z)\sigma_x\partial_z\overleftarrow{\varphi}_{\sigma}(z)-(\partial_z\overleftarrow{\varphi}_{\sigma}^{\dagger}(z))\sigma_x\overleftarrow{\varphi}_{\sigma}(z)\right)
\end{eqnarray}
are the spectral spin currents corresponding to the states incoming from the left (L) and from the right (R), respectively.
Using the wave functions of Eq.~(\ref{states}) we find the spectral spin current in the left/right half space ($z\lessgtr0$):
 \begin{eqnarray}
 \nonumber
 j_{z,L}^{x\sigma}(z<0)&=&\frac{k}{4m}
\left[\chi_{\sigma}^{\dagger}\sigma_x\chi_{\sigma}-\chi_{\sigma}^{\dagger}\hat{r}^{\dagger}\sigma_x\hat{r}\chi_{\sigma}\right]\\
&=&\sigma\frac{k}{2m}\left(1-|r_0|^2-|r_x|^{2}+|r_y|^{2}\right)\\
j_{z,L}^{x\sigma}(z>0)&=&\sigma\frac{k'}{2m}\left(|t_0|^{2}+|t_x|^{2}-|t_y|^{2}\right),
\end{eqnarray}
where we used the identities Eqs.(\ref{r0},\ref{t0}). It is easy to see that the R components of the currents (those corresponding to the states incoming from the right) are obtained by interchanging the half spaces and replacing $k\to k'$, $k'\to k$, and $\hat{r},\;\hat{t}\to\hat{r}',\;\hat{t}'$.



It is worth noting that the spin currents for opposite spin projections are opposite, $j_{z}^{x+}=-j_{z}^{x-}$. Using this fact and setting  $\mu_{R+}= \mu_{R-}$ we rewrite Eq.(\ref{J1}) in the following form
\begin{equation}
\label{J3}
J_z^x=\sum_{{\bf p},k,\sigma}\left[f_F\left(\epsilon_{k,p}-\mu_{L+}\right)-f_F\left(\epsilon_{k,p}-\mu_{L-}\right)\right]j_{zL}^{x+}.
\end{equation}

In the following we consider the zero temperature limit, when the Fermi function is $f_F(E)=\theta(-E)$, and assume that the spin bias $\mu_{L+}-\mu_{L-}=\Delta\mu_L^x$ is much smaller than the Fermi energy, $\Delta\mu_L^x\ll E_F$.
Under these assumptions we can expand the difference of the distribution functions in Eq.(\ref{J3}) to the leading order in $\Delta\mu_L^x$ and simplify the expression for the total current as follows
\be
\label{J4}
J_z^x=\sum_{{\bf p},k}j_{zL}^{x+}\delta(\epsilon_{p,k}-E_F)\Delta\mu_L^x
\ee
which reduces to the following explicit form
\begin{eqnarray}
J_z^{x}(z>0)&=&\sum_{{\bf p},k}\frac{k'}{2m}|t_0|^2\delta(\epsilon_{p,k}-E_F)\Delta\mu_L^x\\
\nn
J_z^{x}(z<0)&=&\sum_{{\bf p},k}\frac{k}{2m}(1-|r_0|^2)\delta(\epsilon_{p,k}-E_F)\Delta\mu_L^x,
\label{jyz}
\end{eqnarray}

\begin{figure}
\begin{center}
\includegraphics[width=3.5in]{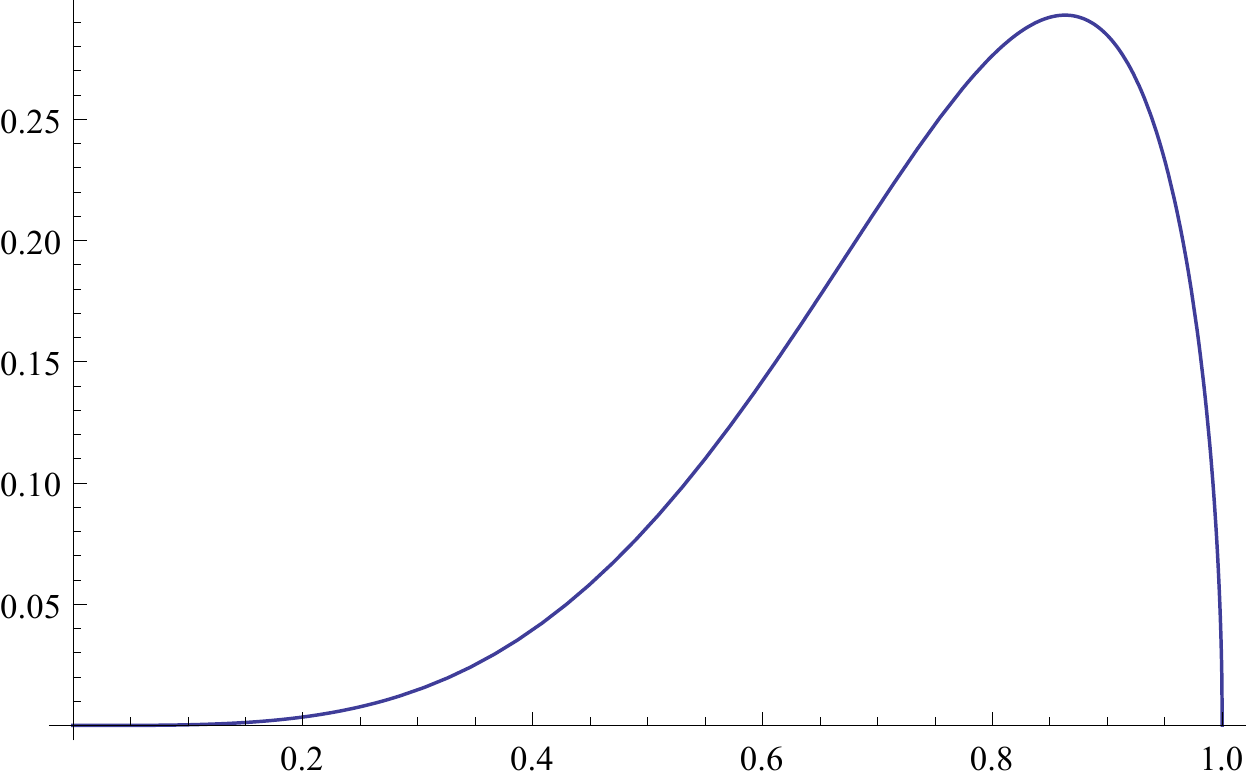}
\caption{ The spin loss $L_s$ in units of the incoming spin current $J_z^{x}(z<0)$, as a function of $\nu$. As we may expect physically it vanishes for $\nu=0,1$.}
\label{SpinLoss}
\end{center}
\end{figure}

Because of ISOC the spin current is not conserved across the interface. The spin loss which occurs at the interface is defined as a discontinuity of the spin current,
\begin{equation}
\label{sl-def}
L^s=J_z^{x}(z<0)-J_z^{x}(z>0),
\end{equation}
which using the condition of Eq.(\ref{current-conserv}) and Eq.~(\ref{Scat2}) can be represented as
\ber
\nn
L_s&=&\sum_{{\bf p},k}\left(\frac{k}{2m}{\bf r}^*{\bf r}+\frac{k'}{2m}{\bf t}^*{\bf t}\right)
\delta(\epsilon_{p,k}-E_F)\Delta\mu_L^x\\
\label{ls}
&=&\frac{s^2\nu^4}{2m}\sum_{{\bf p},k}\frac{p^2|t_0|^2}{k+k'}\delta(\epsilon_{p,k}-E_F)\Delta\mu_L^x.
\eer
Apparently the leading contribution to $L_s$ is of the second order in the dimensionless spin-orbit coupling strength $s$. This agrees with the spin-loss calculated  using the Kubo
formula to explain spin-pumping experiments in the presence of spin-orbit active interfaces \cite{Chen_pumping_2015}. To demonstrate the dependence of the spin-loss Eq.(\ref{ls}) on the difference $V$ of the work functions
we neglect the modification of the transmission coefficient $t_0$ by SOC because for $s<1$ the effects beyond $s^2$ are practically irrelevant. In Fig.(\ref{SpinLoss}) we show the $\nu$-dependence of $L_s$ in units of $J_y^zs^2$. As we can see the spin-loss vanishes for $\nu=0$ because there is no SOC in the absence of the barrier. It also vanishes at $\nu=1$ as at this value the barrier the right half space becomes insulating implying vanishing spin current and no possible spin loss. The spin-loss acquires a maximum near this value due to its $\nu^4$ dependence.

\section{Spin-to-charge conversion}
\label{sec_StC}
In this Section we will calculate the in-plane interface charge current generated by the spin bias -- difference of
spin chemical potentials (x-polarized) across the interface. We will start with the same configuration of the spin chemical potentials as considered in the previous Section, i.e.  $\mu_{L+}\neq \mu_{L-}$ and $\mu_{R+}= \mu_{R-}$.
The total charge current  flowing along the y-direction is described by the following formula.
\ber
\label{JC1}
\nn
J_{y}&=&\sum_{{\bf p},k,\sigma}f_F\left(\epsilon_{k,p}-\mu_{L\sigma} \right)j_{y,L}^{\sigma}
+f_F\left( \epsilon_{k,p}-\mu_{R\sigma}\right)j_{yR}^{\sigma},
\eer
with the spectral charge currents
\begin{eqnarray}
\nn
j_{y,L}^{\sigma}(z)=&-&\frac{ie}{2m}\left(\overrightarrow{\varphi}_{\sigma}^{\dagger}(z)\partial_{y}
\overrightarrow{\varphi}_{\sigma}(z)-(\partial_{y}\overrightarrow{\varphi}_{\sigma}^{\dagger}(z))
\overrightarrow{\varphi}_{\sigma}(z)\right)\\
&-&\frac{e}{m}\left(\overrightarrow{\varphi}_{\sigma}^{\dagger}(z)\hat{\Gamma}_{y}
\overrightarrow{\varphi}_{\sigma}(z)\right)\\
\nn
j_{y,R}^{\sigma}(z)=&-&\frac{ie}{2m}\left(\overleftarrow{\varphi}_{\sigma}^{\dagger}(z)\partial_{y}
\overleftarrow{\varphi}_{\sigma}(z)-(\partial_{y}\overleftarrow{\varphi}_{\sigma}^{\dagger}(z))
\overleftarrow{\varphi}_{\sigma}(z)\right)\\
&-&\frac{e}{m}\left(\overleftarrow{\varphi}_{\sigma}^{\dagger}(z)\hat{\Gamma}_{y}
\overleftarrow{\varphi}_{\sigma}(z)\right),
\end{eqnarray}
where  $\hat{\Gamma}_{y}=\gamma mV'(z)\sigma_{x}$ is the anomalous velocity operator that occurs due to the presence of the ISOC. Accordingly in the currents we will distinguish the ''regular'' or ''normal'' contribution labeled by the superscript $N$, and the "anomalous" one coming from the anomalous velocity, which we will label by the superscript $A$.

The normal contribution to the spectral current reads
\begin{eqnarray}
 \nonumber
 &&j_{y,L}^{N\sigma}(z<0)=\frac{ep_{y}}{2m}\left[2-\chi_{\sigma}^{\dagger}(\hat{r}^{\dagger}e^{2ikz}+
 \hat{r}e^{-2ikz}+\hat{r}^{\dagger}\hat{r})\chi_{\sigma}\right],\\
 &&j_{y,L}^{N\sigma}(z>0)=\frac{ep_{y}}{2m}\chi_{\sigma}^{\dagger}\hat{t}^{\dagger}\hat{t}\chi_{\sigma}
\end{eqnarray}
and similarly for $j_{y,R}^{N\sigma}(z)$. Because of the integration over the directions of ${\bf p}$ in Eq.(\ref{JC1}) only even in $p_y$ or $p_x$ part of the spectral current will contribute to the physical charge current. According to Eqs.(\ref{r0}) the nonvanishing contribution may come only from the terms which are odd in $r_x=t_x\sim p_y$. Therefore the relevant part of the normal spectral current takes the form
\begin{eqnarray}
j_{y,L}^{N\sigma}(z<0)&=&-2\sigma i\frac{e}{m}p_{y}r_x\sin(2kz)\\
j_{y,L}^{N\sigma}(z>0)&=&0.
\end{eqnarray}
As we may see $j_{y,L/R}^{N\sigma}$ is proportional to $\sigma$ which allows us to rewrite Eq.(\ref{JC1}) as follows
\begin{equation}
J_{y}^N(z)=\sum_{{\bf p},k}j_{y,L}^{N+}(z)\delta(\epsilon_{p,k}-E_F)\Delta\mu_L^x,
\end{equation}
where we have used the fact that $\mu_{R+}= \mu_{R-}$ and that $\mu_{L+}-\mu_{L-}=\Delta\mu_L^x\ll E_F$.
Explicitly this equation reads
\begin{eqnarray}
\nonumber
J_{y}^{N}(z<0)&=&-\sum_{{\bf p},k}2i\frac{e}{m}p_{y}r_x\sin(2kz)\delta(\epsilon_{p,k}-E_F)\Delta\mu_L^x\\
&=&-\frac{ek_F^2s}{(2\pi)^2}g(\nu ,2k_Fz)\Delta\mu_L^x\\
J_{y}^{N}(z>0)&=&0.
\end{eqnarray}
\begin{figure}
\begin{center}
\includegraphics[width=3.3in]{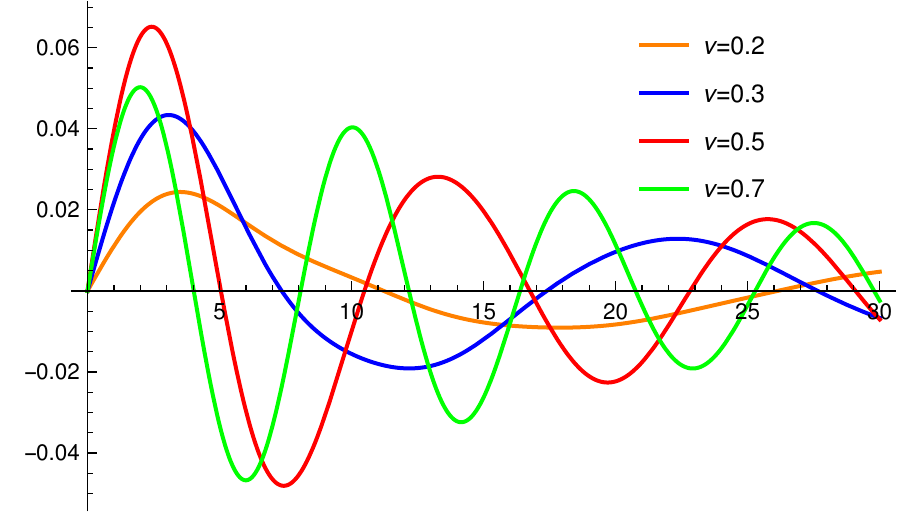}
\caption{The function $g$ as a function of $\nu$ and $z$ in units of $2k_F^{-1}$.
The induced charge current is concentrated near the interface and shows 1D-Friedel oscillations. }
\label{friedel}
\end{center}
\end{figure}

In Fig.(\ref{friedel}) we plot the integral $g(\nu ,2k_Fz)$ at different values of $\nu$ as a function of the distance $z$ from the interface. The charge current is nonzero only in the left metal. It presents clear 1D-Friedel oscillations and a power law decay away from the interface. The normal total contribution to the interface charge current is
\ber
\nonumber
I_{y}^{N}&=&\int dzJ_{y}^{N}(z)\\
&=&\sum_{{\bf p},k}i\frac{e}{km}p_{y}r_x\delta(\epsilon_{p,k}-E_F)\Delta\mu_L^x.
\eer

Let us now calculate the anomalous contribution to the interface charge current.
The corresponding spectral current density reads
\be
j_{y,L}^{A\sigma}=-e\sigma\gamma\left(\overleftarrow{\varphi}_{\sigma}^{\dagger}(z) V'(z)
\overleftarrow{\varphi}_{\sigma}(z)\right)
\ee
The total anomalous current is given by the following expression
\begin{equation}
\label{Jas}
I_{y}^A=-e\gamma\sum_{{\bf p},k}\int dz\overleftarrow{\varphi}^{\dagger}(z) V'(z)
\overleftarrow{\varphi}(z)\delta(\epsilon_{p,k}-E_F)\Delta\mu_L^x
\end{equation}
To obtain the leading in SOC contribution to the anomalous current it is sufficient to neglects the SOC in the scattering states entering the right hand side of Eq.~(\ref{Jas}). At this point we recall the static force balance equation which holds true for any stationary eigenstate the Hamiltonian (\ref{H1}) to the zeroth order in SOC
\be
F_z^{kin}+\varphi^{\dagger}\frac{\partial V}{\partial z}\varphi=0,
\ee
where the kinetic stress stress force is defined as follows (see, for example, Ref.~\onlinecite{Tokatly2005})
\be
F_z^{kin}=\frac{1}{m}\frac{\partial}{\partial z}\left[\frac{\partial\varphi^{\dagger}}{\partial z}
\frac{\partial\varphi}{\partial z}
-\frac{1}{4}\frac{\partial^2}{\partial z^2}(\varphi^{\dagger}\varphi)\right].
\ee
The force balance identity implies for the $z$-integral in Eq.~(\ref{Jas})
\ber
\nn
\int&&dz\overleftarrow{\varphi}^{\dagger}(z) V'(z)\overleftarrow{\varphi}(z)=-\int dz F_z^{kin}\\
&=&-\frac{1}{2m}(k'^2-k^2)t_0^{\dagger}t_0=Vt_0^{\dagger}t_0
\eer
This equation relates the anomalous current to the difference of ``spectral pressures'' at $\pm\infty$,
which only depends on the height $V$ of the potential step, but not of its $z$-dependence.
Using this relation we reduce Eq.(\ref{Jas}) to the following form
\be
\label{I-A}
I_{y}^A=-es\nu^2\sum_{{\bf p},k}\frac{t_0^{*}t_0}{2m}\delta(\epsilon_{p,k}-E_F)\Delta\mu_L^x.
\ee

Now we are able to compare the anomalous and the normal contributions to the current. Both terms
have different signs and a different $\nu$ dependence. The total charge current flowing along the spin-orbit active interface reads
\be
\label{totalc}
I_{y}=I_{y}^A+I_{y}^N=\frac{ek_Fs}{(2\pi)^2}h(\nu)\Delta\mu_L^x.
\ee
\begin{figure}
\begin{center}
\includegraphics[width=3.1in]{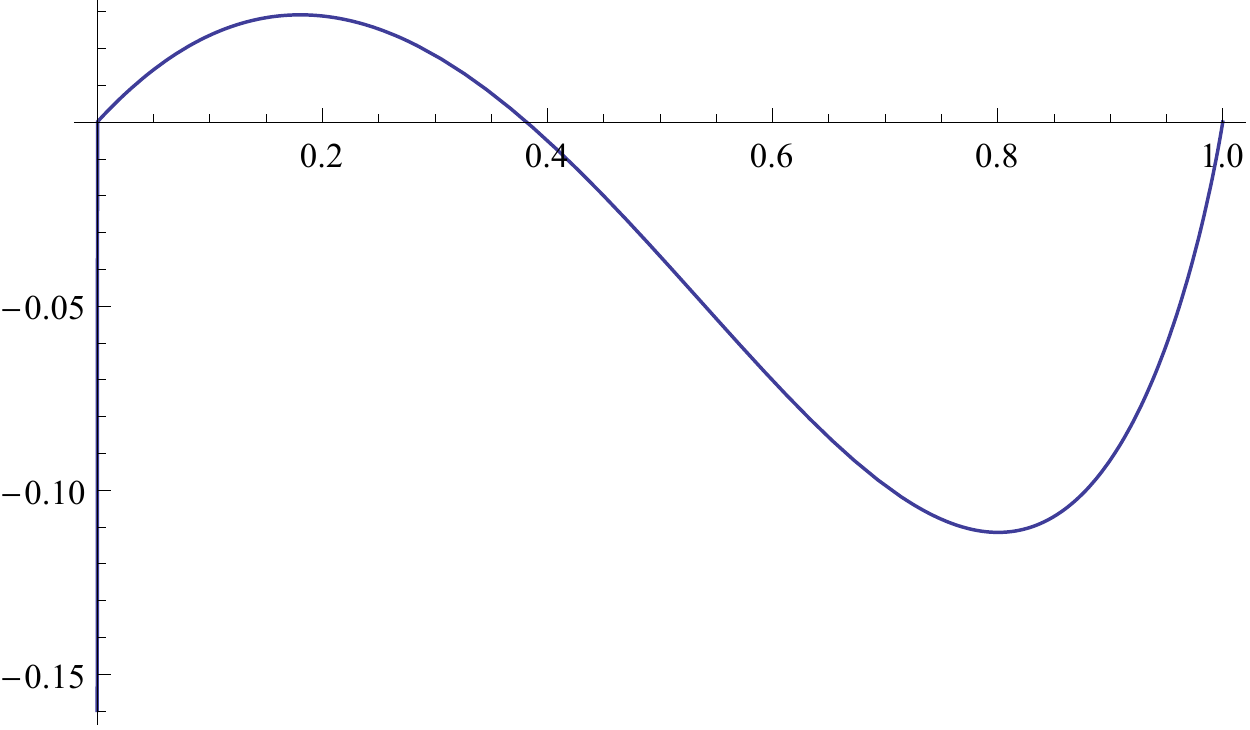}
\caption{The function h($\nu$) which determines the total interface charge current in units of $\frac{ek_Fs}{(2\pi)^2}$. The normal term dominates for small values of $\nu$, implying positive h, while at bigger values of $\nu$ the anomalous term wins giving rise to the sign change. }
\label{totalcurrent}
\end{center}
\end{figure}
The two terms in the current have a different physical origin. The normal contribution to the current originates from the interference of the incoming wave and the wave reflected (with a spin-flip) by the spin-dependent interface. The anomalous contribution has its origin in
the momentum dependence of the SOC term of the Hamiltonian. This term generates the anomalous velocity which is completely localized at the interface.

The different physical nature of the two contributions explains their different $z$- and $\nu$-dependence.
As we can see in Fig.(\ref{totalcurrent}) the difference in the  $\nu$-dependence of the two contributions makes one
term dominate against the other, giving rise to a sign change of the current. The anomalous term depends more strongly on the potential $V$ because it scales with $t_0^2$, while the interference contribution only scales with $t_0$. For this reasons if the barrier is small, the interference term dominates against the one localized at the interface. For larger potentials the anomalous contribution becomes dominant.  As in the case of the spin loss either contribution vanishes at $\nu=0,1$ due to the absence of SOC or because of the insulating nature of the right half space.

We have described the interface charge current produced by imposing the difference of spin-chemical potentials $\Delta\mu_L^x$ on the
left side of the junction, while keeping $\mu_{R+}=\mu_{R-}$ on the right hand side. Obviously the reversed situation with $\Delta\mu_R^x=\mu_{R+}-\mu_{R-}\ne 0$ and $\mu_{L+}=\mu_{L-}$ can be described in a similar way. 
Following the same procedure we obtain the following results for the normal contribution to the current density
\begin{eqnarray}
\nonumber
J_{y}^{N}(z>0)&=&-\sum_{{\bf p},k}2i\frac{e}{m}p_{y}r'_x\sin(2k'z)\\
&\times&\delta(\epsilon_{p,k}-E_F)\Delta\mu_R^x\\
J_{y}^{N}(z<0)&=&0,
\end{eqnarray}
As in the previous case, it shows 1D-Friedel oscillations but now in the $z>0$ half space, while vanishing at $z<0$.
The total normal part of the current is
\begin{eqnarray}
\nonumber
I_{y}^{N}&=&\int dzJ_{x,y}^{N}(z)\\
\nn
&=&\sum_{{\bf p},k}i\frac{e}{k'm}p_{y}r'_x\delta(\epsilon_{p,k}-E_F)\Delta\mu_R^x.
\end{eqnarray}
The anomalous contribution is given by the expression that is similar to Eq.(\ref{I-A})
\be
I_{y}^A=-es\nu^2\sum_{{\bf p},k}\frac{t_0^{'*}t'_0}{2m}\delta(\epsilon_{p,k}-E_F)\Delta\mu_R^x.
\ee

\section{The Spin Swapping effect}
\label{sec_Swappin}
Any linear in momentum SOC can be conveniently described in terms of an effective SU(2) gauge field \cite{Tokatly_Color_PRL08,Gorini10}
$\Acal_i =\Acal_i^a\sigma^a/2$. In our system defined by the Hamiltonian of Eq.(\ref{H1}) the only nonzero components of the SU(2) potential are ${\cal A}_x^y=-{\cal A}_y^x=-\gamma2mV\delta(z)$. The gauge field interpretation of SOC naturally leads to a covariant conservation law for the spin density, which in a steady state requires a vanishing covariant divergence of the spin current ${\cal J}_k={J}_k^a\sigma^a$
\begin{equation}
 \partial_k{\cal J}_k -i [\Acal_k,{\cal J}_k]=0.
\end{equation}
Here the commutator part of the covariant divergence describes the spin torque induced by the SOC.  

In the present case the above covariant continuity equation takes the form
\begin{eqnarray}
\partial_z{\cal J}_z = i\left[{\cal A}_x^y\frac{\sigma^y}{2},{\cal J}_x\right]
-i\left[{\cal A}_y^x\frac{\sigma^x}{2},{\cal J}_y\right],
\end{eqnarray}
which after multiplying by $\sigma^x$ and taking the trace gives
\begin{equation}
\partial_z J_z^x=-\gamma2mV\delta(z)\left(J_x^z+J_y^z\right).
\end{equation}
Integration of this equation across the interface leads the following relation for the spin-loss of Eq.~(\ref{sl-def})
\begin{equation}
\label{continuity}
L_s=s\nu^2\left(J_x^z(0)+J_y^z(0)\right).
\end{equation}
It is worth noting that this general identity does not depend on the validity of the ballistic approximation used in this paper.

The relation of Eq.~(\ref{continuity}) shows that in the presence of the spin-loss generated by ISOC there must exist a nonzero secondary spin current with the spin component perpendicular to that lost at the interface. The symmetry of the problem implies that the spin polarization of the secondary current must be along $\hat{z}$. The reason is that the SOC term Eq.(\ref{H1}) contains only
$\sigma_{x,y}$, so that the primary spin $\sigma_x$ can only rotate about the $\sigma_y$ internal magnetic field,
creating spins polarized in the $\hat{z}$ direction. Therefore only $J_x^z$ is nonzero in the right hand side of Eq.~(\ref{continuity}) meaning that the space and spin indexes of the secondary current are swapped with respect to those of the incident spin current. A similar ``spin current swapping'' effect in which a primary spin current $J_a^{i,P}\equiv J_z^x$ generates a secondary spin current $J_i^{a,SCS}\equiv J_i^z$ is known in the bulk systems with extrinsic SOC \cite{Lifshits2009}. However we are not aware of any discussion of the spin swapping induced by the ISOC. 

To calculate swapped current in the ballistic approximation we follow the procedure used in the previous sections. For the spectral ''secondary`` spin current we find
\begin{eqnarray}
\nn
j_{i,L}^{z\sigma}(z>0)&=&\frac{1}{4m}p_i\chi_{\sigma}^{\dagger}\left[\left(\hat{t}^{\dagger}e^{-ikz}\right)
\sigma_z\left(e^{ikz}\hat{t}\right)\right]\chi_{\sigma}\\
 &=&\sigma\frac{ip_i}{m}\left(t^{*}_yt_0\right)\\
j_{i,L}^{z\sigma}(z<0)&=&\sigma\frac{ip_i}{m}\left(r^{*}_yr_0+r^*_ycos(2kz)\right).
\end{eqnarray}
Now it is easy to see that because of the symmetries of $t_y$ and $r_y$ only the terms with $i=x$ will give a non-zero
contribution after the integration over $\hat{p}$. Therefore we conclude that $J_y^z=0$ and only $J_x^{z}$ can be nonzero.

Following the procedure used in previous sections we obtain the following result for the secondary spin current in the right half space
\be
J_x^{z}(z>0)=s\nu^2\sum_{{\bf p},k}\frac{p_y^2}{m(k+k')}t_0^*t_0\delta(\epsilon_{p,k}-E_F)\Delta\mu_L^x.
\ee
Using Eq.(\ref{ls}) we may rewrite this expression as follows
\be
J_x^{z}(z>0)=\frac{1}{\nu^2 s}L_s,
\ee
which at $z\to 0$ reduces the general identity of Eq.~(\ref{continuity}). In the left metal the secondary spin current
$J_x^{z}(z<0)$ contains two parts, a contribution showing 1D-Friedel oscillations, and a constant one.
\ber
\nn
J_x^{z}(z<0)&=&-\sum_{{\bf p},k}\frac{ip_x}{m}\left(r^{*}_yr_0+r^*_y\cos(2kz)\right)\\
&&\times\delta(\epsilon_{p,k}-E_F)\Delta\mu_L^x
\eer
One can easily check that the current $J_x^{z}(z)$ defined by the above two expressions is continuous at the interface, $J_x^{z}(0^+)=J_x^{z}(0^-)$. The oscillatory part of $J_x^{z}(z<0)$ vanishes upon the $z$-integration (due to the $\cos(2kz)$ dependence instead of $\sin(2kz)$ in the charge current). Therefore only the constant part contributes to the net secondary spin current, which take the following form
\be
J_x^{z<}=s\nu^2\sum_{{\bf p},k}\frac{p_y^2}{m(k+k')}t_0^*r_0\delta(\epsilon_{p,k}-E_F)\Delta\mu_L^x.
\ee

Thus our analysis demonstrates the existence of the global spin current swapping induced by the ISOC. The spin bias across the spin-orbit active interface not only generates a ''trivial`` incident spin current polarized in the bias direction, but also produces the ''swapped`` global spin current that flows along the interface and carries a spin perpendicular to polarization of the spin bias.

\section{Conclusions}
\label{sec_conclusions}

We have developed a simple ballistic model which describes the coupled spin and charge transport in the presence of the spin-orbit coupling generated at the interface between two different metals. This model accounts for different work functions parametrized
by $V$ and the ISOC controlled by the material-dependent parameter $\gamma$. We have
applied the concept of a spin-dependent chemical potential at both sides of the interface in order to create
a non-equilibrium spin density. This model explains qualitatively a number of spin transport phenomena occurring in lateral spin valve geometry and induced by ISOC. In particular we have derived an analytical expression for this spin-loss which describes the reduction of the spin current when it crosses the interface with SOC. We have also calculated the lateral interface charge
current created by the transverse spin bias applied across the junction. Our analysis demonstrates a highly nonmonotonic dependence of this charge current on the difference of work functions. The origin of this dependence is a competition of two contribution to the interface current. The first contribution comes from the
interference between the incoming and the spin-dependent scattered wave functions. The other one has its origin
in the anomalous velocity related to the momentum dependence of the ISOC.
For this reason the first term penetrates the metal and decays a 1D-Friedel oscillations while the
other is completely localized at the interface. Both terms have
different signs and the predominance of one of them depends on the value of the potential $V$. 

Probably the most interesting result of this work is a prediction of the spin current swapping at the interface. We have shown that the bulk lateral ''swapped`` spin current is generated by the spin dependent bias applied across the junction. We have seen that this current is intimately related to the spin loss at the interface. While our specific calculations have been done in the ballistic limit, we believe that the spin current swapping induced by ISOC is a general effect, which can be observed experimentally.

\section{acknowledgements}
We acknowledge financial support from the European Research Council (ERC-2015-AdG-694097),
Spanish  grant  (FIS2016-79464-P),  Grupos  Consolidados  (IT578-13),  AFOSR  Grant  No.
FA2386-15-1-0006 AOARD 144088, H2020-NMP-2014 project
MOSTOPHOS (GA No.   646259) and COST Action MP1306 (EUSpec). J.B. acknowledges  funding from
the  European Union’s Horizon 2020 research and innovation programme under the Marie Sklodowska-Curie
Grant Agreement No. 703195.
\bibliography{refs_SJ-3}
\end{document}